\documentclass[aps,pre,twocolumn,groupedaddress]{revtex4}

\usepackage{graphicx}
\usepackage{amsmath}

\begin{document}

\title{Locations of multicritical points for spin glasses on regular lattices%
}
\author{Masayuki Ohzeki}
\affiliation{Department of Physics, Tokyo Institute of Technology, Oh-okayama, Meguro-ku,
Tokyo 152-8551, Japan}
\date{\today}

\begin{abstract}
We present an analysis leading to precise locations of the
multicritical points for spin glasses on regular lattices. The conventional
technique for determination of the location of the multicritical point was
previously derived using a hypothesis emerging from duality and the replica
method. In the present study, we propose a systematic technique, by an improved
technique, giving more precise locations of the multicritical points on the
square, triangular, and hexagonal lattices by carefully examining
relationship between two partition functions related with each other by the
duality. We can find that the multicritical points of the $\pm J$ Ising
model are located at $p_c = 0.890813$ on the square lattice, where $p_c$
means the probability of $J_{ij} = J(>0)$, at $p_c = 0.835985$ on the
triangular lattice, and at $p_c = 0.932593$ on the hexagonal lattice. These
results are in excellent agreement with recent numerical estimations.
\end{abstract}

\maketitle

\section{Introduction}

The realistic world is affected by randomness. Unlike non-random systems,
those with randomness sometimes show rich and complicated phenomena. One of
the interesting issues for such random systems is spin glass.

Many studies mainly by the mean-field analysis have been successful to
elucidate various concepts for understanding spin glasses \cite{SK,Rev1,Rev2}%
. One of the current issues in spin glasses is their nature in finite
dimensions below the upper critical dimension. Unfortunately, for finite
dimensions, we often rely on numerical simulations, because there are few
ways to analytically study spin glasses in finite dimensions. We need long
equilibration times for the numerical simulations for spin glasses and
average over many realizations of random systems to make error bars small
enough. It is thus difficult to give conclusive understanding on nature of
spin glasses in finite dimensions.

To establish reliable analytical theories of spin glasses has been one of
the most challenging problems for years. A part of successful analyses to
elucidate properties on spin glasses is by the use of the gauge symmetry. By
use of the gauge symmetry, one can obtain the exact value of the internal
energy, evaluate the upper bound for the specific heat, and obtain some
correlation inequalities in a subspace known as the Nishimori line \cite%
{HN81,Rev3}. This gauge symmetry also enables us to rewrite the free energy
along the Nishimori line as the entropy for the distribution of frustration 
\cite{HN86}. Many aspects on spin glasses are essentially related with
frustration. Therefore we expect the possibility that a basis
of establishment of a systematic approach to spin glasses would be in the
gauge symmetry.

\begin{figure}[tbp]
\begin{center}
\includegraphics[width=75mm]{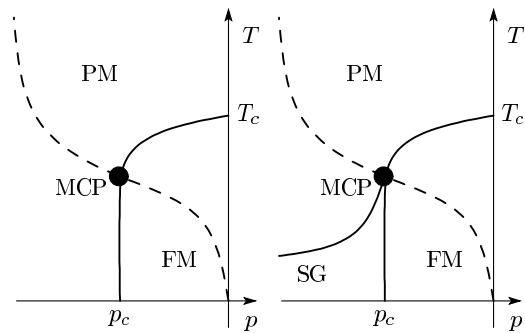}
\end{center}
\caption{{\protect\small Phase diagram of the $\pm J$ Ising model on
two-dimensional lattice (left panel) and on higher dimensions (right panel).
The vertical axis expresses the temperature $T$, and the horizontal line
denotes the concentration $p$ of the antiferromagnetic interactions. The
multicritical point is described by the black point (MCP). The Nishimori
line is described by the dashed line. For higher dimensions, not only the
ferromagnetic (FM) and paramagnetic phases (PM) but also the spin glass
phase (SG) exists.}}
\label{fig1}
\end{figure}

A recent related development with the gauge symmetry is the \textit{%
conjecture} to predict the location of the multicritical point, which is the
special point lying on the intersection between phase boundaries and the
Nishimori line as in Fig. \ref{fig1} \cite{NN,MNN,TN,TSN,NO,NStat}. 
The predictions by the conjecture have shown agreement with numerical
estimations roughly with precision to the third digit. The conjecture has
opened a way of a general scheme for determination of the location of the
multicritical points for spin glasses on any self-dual lattices and mutually
dual pairs of lattices. Nevertheless it has been found that the conjecture
on several mutually dual pairs of hierarchical lattices does not always give
predictions in agreement with estimations by the renormalization group
analysis \cite{HB}. Such discrepancies are not negligible, because the
renormalization group analysis on hierarchical lattices gives exact
solutions. To construct a more reliable technique, we have improved the
technique leading to the location of the multicritical point by combining the concepts of the renormalization group analysis
with the duality \cite{ONB}. The improvement for hierarchical lattices has
greatly succeeded as seen in the literature, because the discrepancies
between the predictions by the improved technique and the exact estimations
by the renormalization group analysis actually decrease. Reconsidering the
improved technique on the hierarchical lattices, in this paper, we apply
the improved technique to the regular lattices such as the square,
triangular, and hexagonal lattices.

The present paper is organized as follows. In Sec. II, we introduce the
conventional conjecture and take a look at several published predictions. In
addition, the problem on the conjecture is pointed out here. The improved
technique on the regular lattices is proposed after review of the case on
the hierarchical lattices in Sec. III, and formulated in Sec. IV. The
improved version shows a very close relationship with the entropy of the
distribution of frustration as shown in this section. In Sec. V, we carry
out the explicit calculations by the improved technique for the regular
lattices. Moreover we have to carefully evaluate the performance of the
improved technique, comparing their predictions with the existing results.
We examine the correspondence with the Domany's exact result \cite%
{Domany1,Domany2} of the slope of the critical point on the phase diagram.
The conclusion is in the last section of the present paper, Sec. VII.

\section{Conventional Conjecture}

\subsection{Duality}

It will be useful to review the analysis by the duality, for the conjecture
is established by combination of the duality and the replica method \cite%
{NN,MNN}. The duality is one of the tools to identify the transition points
for various types of classical spin systems such as the Ising model, and the
Potts model by use of a symmetry embedded in the partition function \cite{KW}%
. We take the non-random Ising model on the square lattice as an example.
The partition function is given as 
\begin{equation}
Z(\beta) = \sum_{\{S_i\}}\prod_{\langle ij \rangle}\exp(\beta J S_i S_j)
\end{equation}
Where $S_i$ is the Ising spin taking $\pm 1$ and the product with the
subscript $\langle ij \rangle$ is over the nearest neighboring sites. We can
regard this partition function as the multi-variable function of components
of the edge Boltzmann factor. In this case, the edge Boltzmann factor is $%
\exp\left(\beta J S_iS_j\right)$). We consider that the dependence on $\beta$
emerges through this edge Boltzmann factors. We set two components as $%
x_0(\beta) = \mathrm{e}^{\beta J}$ and $x_1(\beta) = \mathrm{e}^{-\beta J}$
for convenience. The component $x_0(\beta)$ is often called the principal
Boltzmann factor, which is defined by the edge Boltzmann factor for the
state with all edge spins parallel The principal Boltzmann factor is an
important component throughout this paper. The duality is carried out by the
Fourier transformation for this edge Boltzmann factor defined on each bond
of the lattice \cite{WuWang}. The two-component Fourier transformation gives
the dual edge Boltzmann factors as, 
\begin{eqnarray}
x_0(\beta)^* &=& \frac{1}{\sqrt{2}}\left(\mathrm{e}^{\beta J} + \mathrm{e}%
^{-\beta J}\right) \\
x_1(\beta)^* &=& \frac{1}{\sqrt{2}}\left(\mathrm{e}^{\beta J} - \mathrm{e}%
^{-\beta J}\right).
\end{eqnarray}
As a result, we establish the relation between the partition functions with
different components as, 
\begin{equation}
Z(x_0,x_1) = Z(x^*_0,x^*_1).
\end{equation}
We here extract two principal Boltzmann factors $x_0$ and $x_0^*$ to measure
the energy from the state with edge spins on each bond being parallel as, 
\begin{equation}
x_0^{N_B}(\beta)z(u_1) = x_0^{*N_B}(\beta)z(u^*_1).
\end{equation}
where $N_B$ stands for the number of bonds, and $u_1$ and $u_1^*$ are called
the relative Boltzmann factors defined as $u_1(\beta) = x_1(\beta)/x_0(\beta)
$ and $u^*_1(\beta) = x^*_1(\beta)/x^*_0(\beta)$. Each partition function is
now reduced to a single-variable function of $u_1$ and $u_1^*$, whose
explicit forms are, 
\begin{eqnarray}
u_1(\beta) &=& \mathrm{e}^{-2\beta J} \\
u^*_1(\beta) &=& \tanh \beta J.
\end{eqnarray}
Being very well known, the duality relation can be given as $\mathrm{e}%
^{-2\beta^* J} = \tanh \beta J$ to think of the dual partition function as
one of another Ising model with the edge Boltzmann factor $\mathrm{e}%
^{\beta^* J\sigma_i \sigma_j}$. We can identify the critical point as a
fixed point of the duality as $\mathrm{e}^{-2\beta_c} = \tanh \beta_c J$,
under the assumption of a unique transition. On this critical point, an
appealing equation is satisfied $x_0 = x_0^*$.

\subsection{Duality for the quenched system}

We consider the case of the random-bond Ising model and review the
conventional conjecture \cite{NN,MNN}. The Hamiltonian is defined by 
\begin{equation}
H = - \sum_{\langle ij \rangle} J_{ij} S_i S_j,
\end{equation}
where $J_{ij}$ denotes the quenched random coupling and the summation is
over the nearest neighboring sites. Though various types of distribution for 
$J_{ij}$ can be considered, we here restrict ourselves to the $\pm J$ Ising
model for convenience. The distribution function for the $\pm J$ Ising model
is given by 
\begin{equation}
P(J_{ij}) = p \delta (J_{ij}-J) + (1-p)\delta (J_{ij}+J) = \frac{\mathrm{e}%
^{\beta_p J_{ij}}}{2\cosh \beta_pJ},
\end{equation}
where $\beta_p$ is defined by $\mathrm{e}^{-2\beta_p J} = (1-p)/p$. The
Nishimori line is given by the condition $\beta = \beta_p$ and is described
by the dashed line in each phase diagram of Fig. \ref{fig1}.

We apply the replica method to the $\pm J$ Ising model on the Nishimori line
as, 
\begin{equation}
Z(\beta_p,\beta_p) = \left[ \sum_{\{S_i\}}\prod^n_{\alpha = 1}\exp(\beta_p
J_{ij} S^{\alpha}_i S^{\alpha}_j) \right]_{\mathrm{av}}
\end{equation}
where $n$ stands for the replica number and the angular brackets denote the
configurational average. We apply the duality argument as reviewed above to
this $n$-replicated $\pm J$ Ising model. The duality gives the relationship
of the partition functions with different components of the edge Boltzmann
factor, 
\begin{eqnarray}
& & Z_n(x_0,x_1,\cdots,x_n)  \notag \\
& & \quad = Z_n(x^*_0,x^*_1,\cdots,x^*_n),  \label{PF}
\end{eqnarray}
where the subscript of $x$ and $x^*$ denotes the number of antiparallel-spin
pairs among the $n$ replicas. Two principal Boltzmann factors $x_0$ and $%
x^*_0$ are given as \cite{NN,MNN}, 
\begin{eqnarray}
x_0(\beta) &=& \left[ \mathrm{e}^{n\beta J_{ij}} \right]_{\mathrm{av}}, \\
x^*_0(\beta) &=& \left[ 2^{\frac{n}{2}} \cosh^n\beta J_{ij} \right]_{\mathrm{%
av}}.
\end{eqnarray}
We extract these principal Boltzmann factors similarly to the case of the
non-random bond Ising model, 
\begin{eqnarray}
& & {x_0(\beta)}^{N_B}z_n(u_1,u_2,\cdots,u_n)  \notag \\
& & \quad = {x^*_0(\beta)}^{N_B}z_n(u^*_1,u^*_2,\cdots,u^*_n).  \label{PF1}
\end{eqnarray}
We remark that the partition function for the $n$-replicated $\pm J$ Ising
model is a multi-variable function of the edge Boltzmann factors yet
differently from the case of the non-random bond Ising model.

\subsection{Conjecture}

\begin{figure}[tbp]
\begin{center}
\includegraphics[width=75mm]{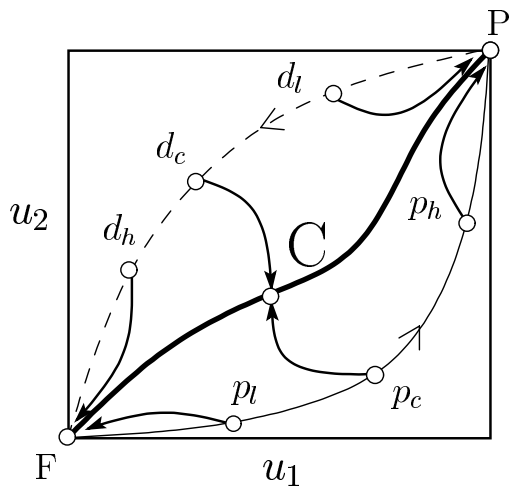}
\end{center}
\caption{{\protect\small Schematic picture of the duality and the
renormalization flow for the replicated $\pm J$ Ising model. }}
\label{Trajectory}
\end{figure}

We describe schematically the relationship two reduced partition functions $%
z_n$ as the curves of the relative Boltzmann factors $(u_1(\beta),u_2(%
\beta),\cdots,u_n(\beta))$ (the thin curve going through the multicritical
point $p_c$) and $(u^*_1(\beta),u^*_2(\beta),\cdots,u^*_n(\beta))$ (the
dashed line) as in Fig. \ref{Trajectory}. We now consider the relationship
between these curves by the projections on the two-dimensional plane $%
(u_1,u_2)$ for convenience. 
As the temperature changes from $0$ to $\infty$, the representative point $%
(u_1(\beta),u_2(\beta),\cdots,u_n(\beta))$ moves toward the point P (the
high-temperature limit) along the thin line in Fig. \ref{Trajectory}. Then
the corresponding dual point $(u^*_1(\beta),u^*_2(\beta),\cdots,u^*_n(\beta))
$ moves along the dashed line in the opposite direction from P to F (the
low-temperature limit). These features have been shown rigorously and imply
the existence of the duality relation for the temperature \cite{NStat}. If
two curves describing change of $(u_1(\beta),u_2(\beta),\cdots,u_n(\beta))$
and $(u^*_1(\beta),u^*_2(\beta),\cdots,u^*_n(\beta))$ become completely
coincident with each other, we can obtain a relation $u_r(\beta^*) =
u^*_r(\beta)$. Solving this relation, we obtain the duality relation for the
temperature $\beta^*(\beta)$. The well-known duality relation $\mathrm{e}%
^{-2\beta^* J} = \tanh \beta J$ for the non-random Ising model can be indeed
derived from the relation $u_r(\beta^*) = u^*_r(\beta)$. Unfortunately the
thin curve $(u_1(\beta),u_2(\beta),\cdots,u_n(\beta))$ does not coincide
with the dashed curve $(u^*_1(\beta),u^*_2(\beta),\cdots,u^*_n(\beta))$ for
the replicated $\pm J$ Ising model on the Nishimori line as shown
schematically in Fig. \ref{Trajectory}. In the case for the replicated $\pm J
$ Ising model, we can neither find a duality relation for the temperature
explicitly, nor identify the multicritical point as the fixed point of
duality.

We thus have provided a hypothesis on determination of the multicritical
point \cite{NN,MNN}. We assume that the equation $x_0(\beta)=x_0^*(\beta)$
is also satisfied at the multicritical point, similarly to the non-random
bond Ising model at the critical point, for any $n$-replicated systems
including the quenched system ($n \to 0$) \cite{NN,MNN,TN,TSN,NO,NStat}.
Validity of this hypothesis can be rigorously shown for $n=1$ and $2$, and
for $n=\infty$, and has been numerically confirmed for the $n=3$ case of the
replicated $\pm J$ Ising model on the square lattice within error bars of
the numerical simulation \cite{MNN}. Relying on these facts, we assume that
we can predict the location of the multicritical point for the $\pm J$ Ising
model on the square lattice by the single equation $x_0(\beta) = x_0^*(\beta)
$. The quenched limit $n \to 0$ for this equation yields \cite{NN,MNN}, 
\begin{equation}
-p \log p - (1-p) \log (1-p) = \frac{1}{2}\log 2.
\end{equation}
The solution to this equation is $p_c = 0.889972$. We can also predict the
location of the multicritical point for other types of randomness. For
instance, that of the Gaussian Ising model with the mean $J_0$ and the
variance $J=1$ is given as $J_0 = 1.021770$ \cite{NN,MNN}. 
\begin{table}[tbp]
\begin{tabular}{lll}
\hline
type & conjecture & numerical result \\ \hline
SQ $\pm J$ & $p_c = 0.889972$\cite{NN,MNN} & $0.8905(5)$ \cite{Aarao} \\ 
&  & $0.8906(2)$\cite{Hone,Picco} \\ 
&  & $0.8907(2)$\cite{Merz} \\ 
&  & $0.8894(9)$\cite{Ito} \\ 
&  & $0.8900(5)$\cite{Queiroz} \\ 
&  & $0.89081(7)$\cite{Hasen} \\ 
SQ Gaussian & $J_0 = 1.021770$ \cite{NN,MNN} & 1.02098(4)\cite{Picco} \\ 
TR $\pm J$ & $p_c = 0.835806$\cite{NO} & $0.8355(5)$\cite{Queiroz} \\ 
TR Gaussian & $J_0 = 0.798174$ & -- \\ 
HEX $\pm J$ & $p_c = 0.932704$\cite{NO} & $0.9325(5)$\cite{Queiroz} \\ 
HEX Gaussian & $J_0 = 1.270615$ & -- \\ \hline
\end{tabular}%
\caption{{\protect\small Comparisons among the results derived by the
conventional conjecture and the existing numerical results. SQ denotes the
square lattice, TR means the triangular lattice, and HEX means the hexagonal
lattice.}}
\label{Comparison}
\end{table}

The conjecture can be extended to mutually dual pair lattices by considering
the product of two partition functions with different couplings indicating
two multicritical points \cite{TSN}, and we can obtain a relation between
these multicritical points. We can predict the location of the multicritical
point on the triangular lattice, though it is the mutually dual pair with
the hexagonal lattice, by simultaneous use of the star-triangle
transformation \cite{NO}. Through these extensions of the conjecture, we
have predicted locations of the multicritical points for various cases,
which show good consistencies with other estimations with high precision in
spite of the simplicity, as seen in Table \ref{Comparison}.

On the other hand, the conjecture has given approximate locations of the
multicritical points on several hierarchical lattices with the largest
discrepancy of $3\% $ \cite{HB}. We have inferred that the issue yielding
such discrepancies comes from the following fact. Unlike the non-random
Ising model, two lines do not coincide as shown in Fig. \ref{Trajectory}.
Nevertheless we have assumed that the equation $x_0 = x_0^*$ is satisfied.
As shown in next section, the technique used in the conjecture has indeed been improved by
reconsideration of these problems.

\section{Improved Technique}

\subsection{Renormalization Flow}

We here introduce a new point of view from renormalization group, which has
helped us to improve the technique on the location of the multicritical point on the hierarchical lattices \cite{ONB}%
. In the present study, we apply this idea to the regular lattices. Two
features through the renormalization group analysis on the hierarchical
lattices are found. (i) The partition function does not change its
functional form by renormalization; only the values of coupling constants
change because of specialty of hierarchical lattices. (ii) The
renormalization flow starting from a critical point is attracted toward an
unstable fixed point. The feature (i) permits us to express the
renormalization flow following the arrows emanating from $p_c$ and $d_c$ to
C, $p_h$ and $d_l$ to P, and $p_l$ and $d_h$ to F as in Fig. \ref{Trajectory}%
. By the feature (ii), there is the renormalization flow from the
multicritical point $p_c$ and corresponding dual point $d_c$, which reaches
the unstable fixed point C, $(u^{(\infty)}_1,u^{(\infty)}_2,\cdots,
u^{(\infty)}_n)$, where the superscripts mean the number of steps of
renormalization. Considering the two properties of the renormalization flow,
we can find that the duality relates two trajectories from $p_c$ and from $%
d_c$, tracing the renormalization flows at each renormalization. In other
words, after a sufficient number of renormalization steps, the thin curve
representing the original system and the dashed curve for the dual system
both approach a common renormalized system represented by the bold line in
Fig. \ref{Trajectory}, which goes through the unstable fixed point C. On
this bold line, the partition function can behave as a single-variable
function and we can identify the unstable fixed point by a single equation $%
x_0(K)=x^*_0(K)$ similarly to the case of the non-random Ising model. This
fact enables us to assume to predict the exact location of the multicritical
point by the following equation, 
\begin{equation}
x^{(\infty)}_0(\beta) = {x_0^*}^{(\infty)}(\beta).  \label{imcon}
\end{equation}
However it is difficult in general to evaluate this equation. We therefore
have proposed a first-approximation equation $x^{(1)}_0={x_0^*}^{(1)}$ as
the improved technique. This equation has indeed led to more precise
results than the relation $x_0(\beta)={x_0^*}(\beta)$ does \cite{ONB}.

\subsection{Partial Summation}

\begin{figure}[tbp]
\begin{center}
\includegraphics[width=75mm]{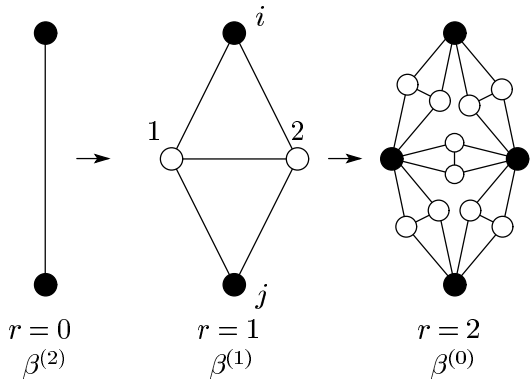}
\end{center}
\caption{{\protect\small A self-dual hierarchical lattice. The number $r$
represents the step of the construction of the hierarchical lattice. The
renormalized inverse temperature is expressed by $\protect\beta^{(l)}$,
where $l$ is the step of the renormalization.}}
\label{fig5}
\end{figure}

We step in the stage of establishment of the improved technique on regular
lattices. Unfortunately, if we apply the above renormalization group
analysis to the regular lattices, the feature (i) may be incorrect, because
other types of many-body interactions are generated after each
renormalization. In addition, the renormalization group analysis for the
regular lattice is usually regarded as an approximate tool, because
many-body interactions appear after the renormalization which prevents us
from iterating the renormalization. If we attempt to construct the recursion
relation of the renormalization for regular lattices, we introduce some
approximations. However we recall that the improved technique on the
hierarchical lattice has been successful within satisfactory precision even
by a one-step renormalization \cite{ONB}. We then sum over internal sites
only in a unit cell of each hierarchical lattice. For example, we show such
a calculation explicitly for the non-random Ising model on a self-dual
hierarchical lattice in Fig. \ref{fig5}. 
\begin{equation}
A\mathrm{e}^{\beta^{(2)}JS_iS_j} = \sum_{S_1,S_2} \mathrm{e}%
^{\beta^{(1)}J\left(S_iS_1 + S_iS_2 + S_{1}S_{2} + S_{1}S_j + S_{2}S_j
\right)},
\end{equation}
where $A$ is the extra coefficient yielded by the renormalization. The
left-hand side corresponds to the principal Boltzmann factor after one-step
renormalization when $S_i S_j = 1$. We can regard this renormalized
principal Boltzmann factor as the partition function defined on the unit
cell of the hierarchical lattice in Fig. \ref{fig5} under the constraint
that the spins at both ends are parallel. In other words, the principal
Boltzmann factor for the improved technique is in general given as 
\begin{eqnarray}
& & Z(\beta) = \overline{\sum_{\{S_i\}}} \prod^{\mathrm{unit}}_{\langle ij
\rangle} \mathrm{e}^{\beta^{(0)}J S_iS_j}
\end{eqnarray}
where the overline means the summation over internal spins $S_1$ and $S_2$
on the unit cell of the hierarchical lattice with the spins on the edge up $%
S_i$ and $S_j = 1$. The product runs over the nearest neighboring pairs on
the unit of the hierarchical lattice. For the case for the random-bond Ising
model, after application of the replica method, we also obtain the similar
principal Boltzmann factor to the above example, by performing the partial
summation over the unit of the hierarchical lattice. Instead of iterating
summation as the renormalization, we sum partially over sites on the regular
lattice. After we partially trace out the degrees of freedom in some area of
the regular lattice, we regard the partition function on this area as the
principal Boltzmann factor as the above example. 
We are able to construct the improved technique for the regular lattice
without neglecting many-body interactions emerging after the summation. We
call the limited-summation area of the regular lattice the cluster in this
paper, and we define the partition function on the cluster as the principal
Boltzmann factor for the improved technique on the regular lattice. When we
take the summation over the internal sites, we impose the fixed boundary
condition on all the spins on the boundary of the cluster as we have done
for edge Boltzmann factor in the conventional conjecture.

We remark the possibility of the performance of the improved technique.
Even if we partially sum over degrees of freedom on the regular lattice as
considered above, it is necessary to use an infinite-dimensional space to
simultaneously express changes and generations of various types of many-body
interactions. In this infinite dimensional space, we assume that there are
two renormalization flows going uniformly toward the unstable fixed point,
similarly to the case on hierarchical lattices. It is not able to examine
such behaviors of the renormalization flows in the infinite-dimensional
space and we cannot verify this assumption. Equation $x_0={x_0^*}$ on the
cluster may hence be a worse approximation than the conventional conjecture $%
x^{(0)}_0={x_0^*}^{(0)}$ by consideration on a single bond. We have to
carefully evaluate the performance of the improved technique for the
regular lattices. At least, the obtained results as seen later can, however,
give an answer for the location of the multicritical point for the $\pm J$
Ising model on the square lattice with the precision to the fourth digit.

In addition, we will consider several types of the improvements for the
regular lattice. On the hierarchical lattices, sufficient renormalization
enables us to obtain the exact solution. Therefore we can expect that the
improved technique for the hierarchical lattices gives more precise answers
if we use the principal Boltzmann factor after more renormalization steps.
On the regular lattices, it is considered that the performance of the
improved technique depends on the area of the cluster. The number of
degrees of freedom partially summed over on the regular lattice would
correspond the step of the renormalization on the hierarchical lattice. We
propose a few types of the clusters including several bonds below,
considering this assumption.

\section{Formulation}

\subsection{Square Lattice}

\begin{figure}[tb]
\begin{center}
\includegraphics[width=75mm]{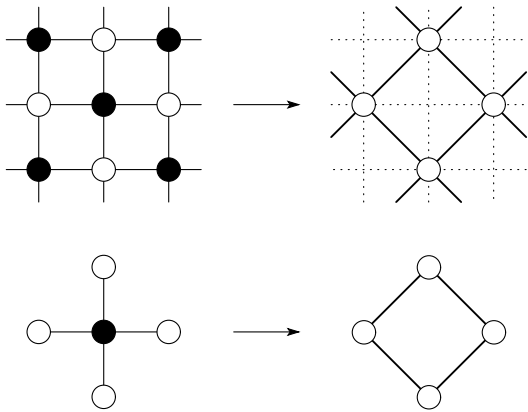}
\end{center}
\caption{{\protect\small Example of the partial summation for the square
lattice and the cluster. The top figures describe one of the types of the
summation for the square lattice. The bottom figures express the cluster for
the principal Boltzmann factor.}}
\label{fig6}
\end{figure}

The starting point for the establishment of the improved technique is the
exact duality relation (\ref{PF}) for the $n$-replicated partition function.
As shown in Fig. \ref{fig6}, we consider to sum over a part of the spins on
the square lattice. 
Then the exact duality relation (\ref{PF}) is reduced to 
\begin{equation}
Z^{(s)}_n(x^{(s)}_0,x^{(s)}_1,\cdots,x^{(s)}_n) =
Z^{(s)}_n(x^{*(s)}_0,x^{*(s)}_1,\cdots,x^{*(s)}_n).
\end{equation}
Here $Z^{(s)}_n$ represents the reduced partition function by the summation
of a part of spins on the square lattice. The superscript $s$ distinguishes
the type of the approximations, for we consider different types of the
summation below. The quantity $x^{(s)}_k$ is the edge Boltzmann factor
including many-body interactions generated after the summation. We take a
cluster of the square lattice as in Fig. \ref{fig6} and define the principal
Boltzmann factors after the summation, 
\begin{eqnarray}
x^{(s)}_0 &=& \left[ \left\{ \overline{\sum_{\{S_i\}}} \prod^{\mathrm{part}%
}_{\langle ij \rangle} \mathrm{e}^{\beta J_{ij}S_i S_j} \right\}^n \right]_{%
\mathrm{av}} \\
x^{*(s)}_0 &=& \left[ \left\{ \overline{\sum_{\{S_i\}}} \prod^{\mathrm{part}%
}_{\langle ij \rangle} \frac{1}{\sqrt{2}}\left(\mathrm{e}^{\beta J_{ij}}+ 
\mathrm{e}^{-\beta J_{ij}} S_i S_j\right) \right\}^n \right]_{\mathrm{av}}, 
\notag \\
& &
\end{eqnarray}
where the overline means the summation over internal spins in the cluster of
the square lattice as the filled circle in Fig. \ref{fig6} with the other
spins fixed to up directions $\{S_i = 1\}$ represented by the white circles
in Fig. \ref{fig6}. The word ``part" represents that the product runs over
the bonds of the cluster under consideration. These principal Boltzmann
factors can be regarded as partition functions after the configurational
average and application of the replica method defined on the cluster under
the fixed boundary condition as shown in Fig. \ref{fig6}.

We assume that a single equation gives the critical points for any number of 
$n$, similarly to the conventional conjecture, 
\begin{equation}
x^{(s)}_0 = x^{*(s)}_0.
\end{equation}
By the extrapolation of the quenched limit $n \to 0$ of this equation, we
establish the improved technique for the square lattice as follows, 
\begin{eqnarray}
& & \left[ \log Z^{*(s)}(\beta,\{J_{ij}\}) \right]_{\mathrm{av}} - \left[
\log Z^{(s)}(\beta,\{J_{ij}\}) \right]_{\mathrm{av}} = 0.  \notag \\
& &  \label{6IC1}
\end{eqnarray}
We need the configurational average for $J_{ij}$ of the logarithmic terms by
two partition functions $Z^{(s)}$ and $Z^{*(s)}$ defined on the cluster, 
\begin{eqnarray}  \label{Appdual}
Z^{(s)}(\beta,\{J_{ij}\}) &=& \overline{\sum_{\{S_i\}}} \prod^{\mathrm{part}%
}_{\langle ij \rangle} \mathrm{e}^{\beta J_{ij} S_i S_j}  \label{App} \\
Z^{*(s)}(\beta,\{J_{ij}\}) &=& \overline{\sum_{\{S_i\}}} \prod^{\mathrm{part}%
}_{\langle ij \rangle} \frac{1}{\sqrt{2}} \left(\mathrm{e}^{\beta J_{ij}}+ 
\mathrm{e}^{-\beta J_{ij}} S_i S_j \right),  \notag \\
\end{eqnarray}
where the asterisk means that the edge Boltzmann factor is given in a
different form obtained after the duality for the $\pm J$ Ising model. We
can estimate the location of the multicritical point by the above relation (%
\ref{6IC1}) as detailed below.

\subsection{Triangular Lattice}

Similarly to the case on the square lattice, we can derive the improved
technique for the triangular lattice. We give first several
remarks on the duality with the star-triangle transformation. For the
triangular lattice, it is convenient to use the face Boltzmann factor
instead of the edge Boltzmann factor. The original face Boltzmann factor has
three edge Boltzmann factors defined on the bonds of the elementary
triangle. On the other hand, the dual face Boltzmann factor has three dual
edge Boltzmann factors defined on the bonds on the star shape as in Fig. \ref%
{fig9} \cite{NO,WuPotts}. In the dual face Boltzmann factor, the summation
over the spin at the center of the star is included, which corresponds to
the star-triangle transformation. 
\begin{figure}[tbp]
\begin{center}
\includegraphics[width=75mm]{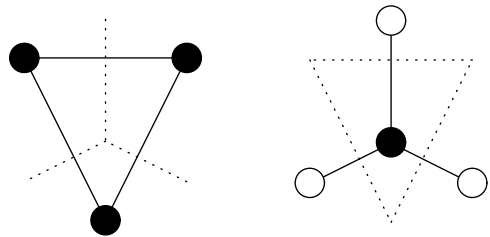}
\end{center}
\caption{{\protect\small Duality for the triangular lattice. After the
duality transformation, the partition function on the triangular lattice is
transformed into another partition function on the hexagonal lattice. }}
\label{fig9}
\end{figure}
Similarly to the case of the square lattice, we start from the exact duality
relation for the replicated random-bond Ising model on the triangular
lattice, 
\begin{equation}
Z_{\mathrm{TR},n}(A_0,A_1,\cdots) = Z^{(s)}_{\mathrm{TR},n}(A^*_0,A^*_1,%
\cdots),
\end{equation}
where $A$ and $A^*$ are the original and dual face Boltzmann factors. We
consider the summation over a part of the spins on the triangular lattice as
in Fig. \ref{TRpart}, and then this equation is reduced to, 
\begin{equation}
Z^{(s)}_{\mathrm{TR},n}(A^{(s)}_0,A^{(s)}_1,\cdots) = Z^{(s)}_{\mathrm{TR}%
,n}(A^{*(s)}_0,A^{*(s)}_1,\cdots),
\end{equation}
where $Z^{(s)}_{\mathrm{TR},n}$ represents the reduced partition function by
the summation over a part of spins on the triangular lattice. 
\begin{figure}[tb]
\begin{center}
\includegraphics[width=75mm]{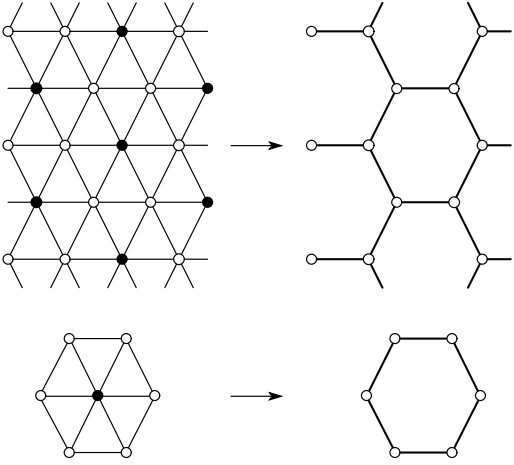}
\end{center}
\caption{{\protect\small Example of the summation for the triangular lattice
and the cluster. The top figures express one of the types of the summation
for the triangular lattice. The bottom figures express the cluster for the
evaluation of the principal Boltzmann factor.}}
\label{TRpart}
\end{figure}
The quantity $A^{(s)}_k$ is the face Boltzmann factor after the summation.
We define the principal Boltzmann factor after the summation over a part of
the triangular lattices as, 
\begin{eqnarray}
A^{(s)}_0 &=& \left[ \left\{ \overline{\sum_{\{S_i\}}} \prod^{\mathrm{part}%
}_{\bigtriangleup} \mathrm{e}^{\beta J_{12}S_1S_2 + \beta J_{23}S_2 S_3 +
\beta J_{31}S_3 S_1} \right\}^n \right]_{\mathrm{av}},  \notag \\
& & \\
A^{*(s)}_0 &=& \left[ 2^{-\frac{n}{2}N_s^{(s)}} \right.  \notag \\
& & \quad \times \left. \left\{ \overline{\sum_{\{S_i\}}}\sum_{\{S_0\}}
\prod^{\mathrm{part}}_{\bigtriangleup} \frac{1}{\sqrt{2}}\left(\mathrm{e}%
^{\beta J_{10}}+ \mathrm{e}^{-\beta J_{10}} S_1 S_0\right) \right.\right. 
\notag \\
& & \qquad \times \left. \left. \frac{1}{\sqrt{2}}\left(\mathrm{e}^{\beta
J_{20}}+ \mathrm{e}^{-\beta J_{20}} S_2 S_0\right) \right. \right.  \notag \\
& & \quad \qquad \times \left. \left. \frac{1}{\sqrt{2}}\left(\mathrm{e}%
^{\beta J_{30}}+ \mathrm{e}^{-\beta J_{30}} S_3 S_0\right) \right\}^n \right]%
_{\mathrm{av}},  \notag \\
\end{eqnarray}
where $N_s^{(s)}$ is equal to the number of the up-pointing triangles
included in the cluster. The overline means the summation over internal
spins included in the cluster of the triangular lattice as the filled
circles in Fig. \ref{TRpart} with the other spins (the white circles) up $%
\{S_i = 1\}$. The word ``part" represents summation over the up-pointing
triangles in the cluster as in Fig. \ref{TRpart}. The summation over $S_0$
means the star-triangle transformation. We can establish the improved
technique for the triangular lattice by an equation of these principal
Boltzmann factors after the summation $A^{(s)}_0 = A^{*(s)}_0$ and by the
extrapolation to the quenched limit of $n \to 0$, 
\begin{equation}
\left[ \log Z_{\mathrm{TR}}^{*(s)}(\beta,\{J_{ij}\}) \right]_{\mathrm{av}} - %
\left[ \log Z_{\mathrm{TR}}^{(s)}(\beta,\{J_{ij}\}) \right]_{\mathrm{av}} =
0,  \label{6ICTR}
\end{equation}
where the partition functions $Z_{\mathrm{TR}}^{*(s)}$ and $Z_{\mathrm{TR}%
}^{*(s)}$ on the cluster of the triangular lattice are defined as, 
\begin{eqnarray}
& & Z_{\mathrm{TR}}^{(s)}(\beta,\{J_{ij}\})  \notag \\
& & \quad = \left[\overline{\sum_{\{S_i\}}} \prod^{\mathrm{part}%
}_{\bigtriangleup} \mathrm{e}^{\beta J_{12}S_1S_2 + \beta J_{23}S_2 S_3 +
\beta J_{31}S_3 S_1} \right]_{\mathrm{av}} \\
& & Z_{\mathrm{TR}}^{*(s)}(\beta,\{J_{ij}\})  \notag \\
& & \quad = \left[ 2^{-\frac{1}{2}N_s^{(s)}} \overline{\sum_{\{S_i\}}}%
\sum_{\{S_0\}} \prod^{\mathrm{part}}_{\bigtriangleup} \frac{1}{\sqrt{2}}%
\left(\mathrm{e}^{\beta J_{10}}+ \mathrm{e}^{-\beta J_{10}} S_1 S_0\right)
\right.  \notag \\
& & \quad \quad \times \left.\frac{1}{\sqrt{2}}\left(\mathrm{e}^{\beta
J_{20}}+ \mathrm{e}^{-\beta J_{20}} S_2 S_0\right)\right.  \notag \\
& & \quad \qquad \times \left.\frac{1}{\sqrt{2}}\left(\mathrm{e}^{\beta
J_{30}}+ \mathrm{e}^{-\beta J_{30}} S_3 S_0\right) \right]_{\mathrm{av}}.
\end{eqnarray}

\subsection{Frustration Entropy and Multicritical Point}

Before going into the detailed calculations for determination of the
locations of the multicritical points, we consider the physical meaning of
the improved technique from a different point of view. We show the
relationship between the improved technique and the gauge symmetry, by
taking the case of the square lattice as an example. The observation below
implies the existence of deeply physical meaning behind the multicritical
point. We will discuss the structure of the phase diagram by use of the
connection between the improved technique and the gauge symmetry as shown
below in the following section.

The second term $Z^{(s)}$ on the left-hand side of Eq. (\ref{6IC1}) can be
regarded as the free energy for the random-bond Ising model defined on the
cluster if divided by $\beta$. The free energy for the gauge invariant model
can be reduced to another form by the gauge transformation defined as \cite%
{HN81,Rev3}, 
\begin{eqnarray}
J_{ij} &\to& J_{ij}\sigma_i \sigma_j  \label{GT1} \\
S_{i} &\to& S_{i}\sigma_i,  \label{GT2}
\end{eqnarray}
where $\sigma_i$ takes either $-1$ or $+1$. We can obtain a useful
expression of $\log Z^{(s)}$ for the case of the $\pm J$ Ising model for
instance \cite{HN86}, 
\begin{eqnarray}
& & \left[\log Z^{(s)}(\beta,\{J_{ij}\})\right]_{\mathrm{av}}  \notag \\
& & \quad = \frac{1}{2^{N^{(s)}_s}(2\cosh \beta_p J)^{N^{(s)}_B}}  \notag \\
& & \qquad \times \sum_{\{J_{ij}\}} Z^{(s)}(\beta_p,\{J_{ij}\}) \log
Z^{(s)}(\beta,\{J_{ij}\}),  \label{Fen}
\end{eqnarray}
where $N_B^{(s)}$ is the number of bonds included in the cluster. In this
expression, we obtain the entropy of the distribution of frustration on the
cluster by setting $\beta= \beta_p$.

\begin{figure}[tb]
\begin{center}
\includegraphics[width=75mm]{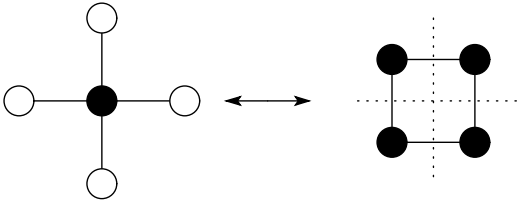}
\end{center}
\caption{{\protect\small Dual lattice for the cluster in Fig. \protect\ref%
{fig6}. The black colored sites are free spins as the targets of the
summation and the white ones are fixed in the up-pointing directions.}}
\label{fig28}
\end{figure}

On the other hand, the first term $Z^{*(s)}$ on the left-hand side of Eq. (%
\ref{6IC1}), which is generated from the dual principal Boltzmann factor
after the summation, is not in a gauge invariant form. The duality, however,
can transform $Z^{*(s)}$ into a gauge invariant form. At this time, the form
of the edge Boltzmann factor changes from $\left\{\exp(\beta J_{ij}) +
\exp(-\beta J_{ij})S_i S_j \right\}/\sqrt{2}$ to $\exp\left(\beta J_{ij}
S_iS_j\right)$ \cite{WuWang}, 
\begin{equation}
Z^{*(s)} = 2^{N^{(s)}_s - \frac{N^{(s)}_B}{2} - 1}Z^{(s)}_{\mathrm{D}}[x],
\label{SSQ2}
\end{equation}
where $N_s^{(s)}$ is the number of sites in the cluster of the original
square lattice. However the form of the lattice is transformed as shown in
Fig. \ref{fig28}. We denote this fact by the subscript D. 
The application of the duality to the partition function $Z^{*(s)}$ on the
cluster with the dual edge Boltzmann factor enables us to derive a gauge
invariant form, 
\begin{eqnarray}
Z_{\mathrm{D}}^{(s)}(\beta,\{J_{ij}\}) &=& \overline{\sum_{\{S_i\}}} \prod^{%
\mathrm{part(D)}}_{\langle ij \rangle} \mathrm{e}^{\beta J_{ij} S_i S_j},
\label{GInZ}
\end{eqnarray}
where the `part(D)' expresses the product over the bonds on the dual cluster
as in Fig. \ref{fig28}. The configurational-averaged quantity of its
logarithm can be regarded as the free energy of the random-bond Ising model
defined on the dual lattice for the small lattice. We use the relation (\ref%
{SSQ2}) and rewrite the first term $Z^{*(s)}$ on the left-hand side of Eq. (%
\ref{6IC1}) as, 
\begin{eqnarray}
& & \left[ \log Z^{*(s)}(\beta,\{J_{ij}\}) \right]_{\mathrm{av}}  \notag \\
& & \quad = \left[ \log Z_{\mathrm{D}}^{(s)}(\beta,\{J_{ij}\}) \right]_{%
\mathrm{av}} + \left(N^{(s)}_s - \frac{N^{(s)}_B}{2} - 1 \right)\log 2. 
\notag \\
\end{eqnarray}
The first term on the right-hand side of this relation can be reduced to the
same expression as Eq. (\ref{Fen}), 
\begin{eqnarray}
& & \left[\log Z_{\mathrm{D}}^{(s)}(\beta,\{J_{ij}\})\right]_{\mathrm{av}} 
\notag \\
& & \quad = \frac{1}{2^{N_{\mathrm{D}}^{(s)}}(2\cosh \beta_p J)^{N^{(s)}_B}}
\notag \\
& & \qquad \times \sum_{\{J_{ij}\}} Z_{\mathrm{D}}^{(s)}(\beta_p,\{J_{ij}\})
\log Z_{\mathrm{D}}^{(s)}(\beta,\{J_{ij}\}).
\end{eqnarray}
Here $N_D^{(s)}$ expresses the number of sites of the dual cluster, which is
equal to that of the plaquettes of the original cluster. In the case of the
cluster of the square lattice as in Fig. \ref{fig28}, $N_s^{(s)} = 1$, $%
N_B^{(s)} = 4$, $N_D^{(s)} = 4$. The above considerations enable us to
rewrite Eq. (\ref{6IC1}) as, 
\begin{eqnarray}
& & \frac{1}{2^{N^{(s)}_D}}S_{\mathrm{D}}^{(s)}(\beta_p,\beta) - \frac{1}{%
2^{N^{(s)}_s}} S^{(s)}(\beta_p,\beta)  \notag \\
& & \quad = \left(\frac{N^{(s)}_B}{2} - N^{(s)}_s + 1 \right) \log2,
\label{6IC2}
\end{eqnarray}
where 
\begin{eqnarray}
S_{\mathrm{D}}^{(s)}(\beta_p,\beta) &=& \sum_{\{J_{ij}\}}\frac{Z_{\mathrm{D}%
}^{(s)}(\beta_p,\{J_{ij}\})}{\left( 2\cosh \beta_p J \right)^{N^{(s)}_B}}
\log \frac{Z_{\mathrm{D}}^{(s)}(\beta,\{J_{ij}\})}{\left( 2\cosh \beta J
\right)^{N^{(s)}_B}},  \notag \\
& & \\
S^{(s)}(\beta_p,\beta) &=& \sum_{\{J_{ij}\}}\frac{Z^{(s)}(\beta_p,\{J_{ij}\})%
}{\left( 2\cosh \beta_p J \right)^{N^{(s)}_B}} \log \frac{%
Z^{(s)}(\beta,\{J_{ij}\})}{\left( 2\cosh \beta J \right)^{N^{(s)}_B}}. 
\notag \\
\end{eqnarray}
If we set $\beta = \beta_p$ in this expression, we find that Eq. (\ref{6IC2}%
) states that the multicritical point is located where the difference
between two entropies of the distribution of frustration on two lattices
related by the duality takes a special value. We can use this expression of
the improved technique to lead the structure of the phase diagram such that
the gauge symmetry and several correlation inequalities used to give the
prediction as in Fig. \ref{fig1} \cite{HN81,Rev3}. It is straightforward to
establish the expression as in Eq. (\ref{6IC2}) also for the case of the
triangular lattice.

In the next section, we show the results for the precise locations of the
multicritical points obtained by computing Eq. (\ref{6IC1}) for the square
lattice and Eq. (\ref{6ICTR}) for the triangular lattice.

\section{Derivations of Multicritical Points}

\begin{figure}[tb]
\begin{center}
\includegraphics[width=75mm]{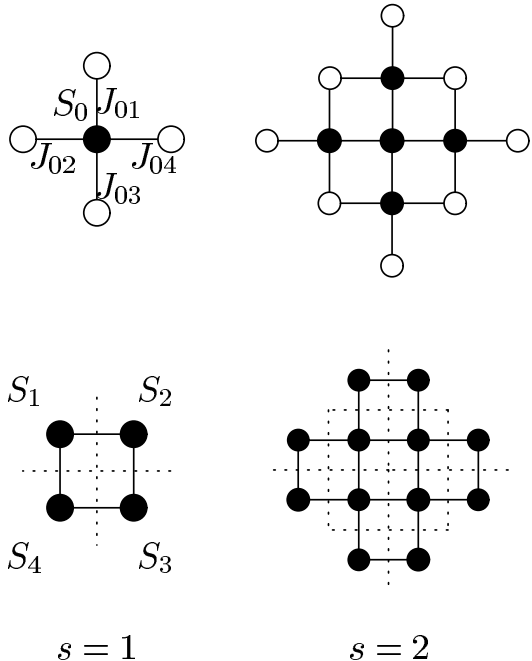}
\end{center}
\caption{{\protect\small Two patterns of the clusters for the improved
technique on the square lattice. The top figures express the clusters for
the evaluations of $Z^{(s)}$ and $Z^{*(s)}$. The bottom figures represent
the dual lattices for the clusters, on which the partition functions are
denoted by $Z^{(s)}_{\mathrm{D}}$. The filled circles are the targets of the
summation and white ones are fixed to up directions $\{S_i\}=1$.}}
\label{fig29r}
\end{figure}

\begin{figure}[tb]
\begin{center}
\includegraphics[width=90mm]{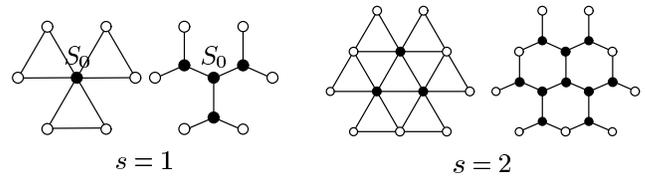}
\end{center}
\caption{{\protect\small Two patterns of the clusters for the improved
technique on the triangular lattice. The left-hand side for each type of
the approximations is for the partition function $Z_{\mathrm{TR}}^{(s)}$,
and the right-hand side is for the dual partition function $Z_{\mathrm{TR}%
}^{*(s)}$, for which the star-triangle transformation is needed. We use the
same symbols as in Fig. \protect\ref{fig29r}. }}
\label{fig30}
\end{figure}

We derive the location of the multicritical point by the improved technique
for the regular lattices. If we consider a larger range of the summation of
spins, (i. e., the cluster includes more bonds and sites.) it is expected
that the precision of the improved technique becomes higher. One of the
reasons is that the improved technique can include more effects of
spatially non-uniform interactions, which are essential features in random
spin systems. In other words, the conventional conjecture has been the
zeroth approximation without consideration of a form of the lattice and
non-uniform interactions in space. The improved technique is also an
approximation but gives more precise answers than the conventional
conjecture, because it is formulated with the consideration of an individual
characteristic of the lattice similarly to the case of the hierarchical
lattices. We express the type of the approximations by the value of $s$,
which has represented the form of the cluster. In this paper, we show the
derivations of the multicritical point by use of several types of the
clusters for the cases on the square lattice as in Fig. \ref{fig29r} and on
the triangular lattice as in Fig. \ref{fig30}. 

\subsection{First Approximation for the Square Lattice}

We consider the first approximation for the location of the multicritical
point on the square lattice by the $s=1$ cluster as shown in Fig. \ref%
{fig29r}. To identify the multicritical point, we solve Eq. (\ref{6IC1}). We
first calculate the partition function on the cluster in Eq. (\ref{App}).
The summation over the single spin $S_0$ at the center surrounded by four
bonds $(J_{01},J_{02},J_{03},J_{04})$ with four spins up yields 
\begin{eqnarray}
& & Z^{(1)}(\beta, \{J_{ij}\})  \notag \\
& & \quad = \sum_{S_0=\pm 1}\mathrm{e}^{%
\beta(J_{01}+J_{02}+J_{03}+J_{04})S_0}  \notag \\
& & \qquad = 2\cosh \{ \beta(J_{01}+J_{02}+J_{03}+J_{04})\}.
\end{eqnarray}
Another partition function in Eq. (\ref{Appdual}) is calculated as, 
\begin{eqnarray}
& & Z^{*(1)}(\beta, \{J_{ij}\})  \notag \\
& & \quad = \left(\frac{1}{\sqrt{2}}\right)^4 \sum_{S_0=\pm 1} \prod_{i=1}^4
\left( \mathrm{e}^{\beta J_{0i}} + S_0\mathrm{e}^{-\beta J_{0i}}\right) 
\notag \\
& & \qquad = \left(\frac{1}{\sqrt{2}}\right)^{4}
\left\{\prod_{i=1}^{4}\left(2\cosh \beta J_{0i}\right) + \prod_{i=1}^{4}
\left(2\sinh \beta J_{0i}\right) \right\}.  \notag \\
\end{eqnarray}
This quantity is also obtained from the evaluation of the partition function 
$Z^{(1)}_{\mathrm{D}}$ defined on the dual cluster by use of the relation (%
\ref{SSQ2}). We can explicitly write down the equation for the precise location of the multicritical point for the $\pm J$ Ising model on the
square lattice as, from Eq. (\ref{6IC1}), 
\begin{eqnarray}
& & \sum_{\tau_{ij}}\frac{1}{2^4} \left(1 + \tanh^4 K_p \prod_{i=1}^4
\tau_{0i} \right) \log \left(1 + \tanh^4 K \prod_{i=1}^4 \tau_{0i} \right) 
\notag \\
& & - \sum_{\tau_{ij}}\frac{1}{2}~\frac{2\cosh\left\{K_p\sum_{i=0}^4%
\tau_{0i}\right\}}{\left(2\cosh K_p\right)^4} \log \frac{2\cosh\left\{K%
\sum_{i=0}^4\tau_{0i}\right\}}{\left(2\cosh K\right)^4}  \notag \\
& & \quad = 2 \log 2,
\end{eqnarray}
where we use the coupling constant $K =\beta J$ and its sign $\tau_{ij}$.
Setting $K = K_p$, we solve this equation and obtain $p^{(1)}_c = 0.890725$.
This result is listed in Table \ref{6ICTable} to see the performance of the
improvement and to compare the results by the improved technique with the existing
ones. Another type of the approximations for the square lattice is by the
cluster labeled by $s=2$ in Fig. \ref{fig29r} and can be straightforwardly
evaluated. The numerical manipulation of this approximation give another
prediction $p^{(2)}_c = 0.890822$ as listed in Table \ref{6ICTable}.

For the Gaussian Ising model, we have to evaluate the quadruple integration
over four bonds $\{J_{ij}\}$ for the improved technique as, 
\begin{widetext}
\begin{eqnarray}\nonumber
& & \int^{\infty}_{-\infty}\prod_{i=1}^{4}P(J_{0i})dJ_{0i} \log \left\{\prod_{i=1}^{4}\left(2\cosh \beta J_{0i}\right) + \prod_{i=1}^{4} \left(2\sinh \beta J_{0i}\right) \right\} \\ \nonumber
& & \quad - \int^{\infty}_{-\infty}\prod_{i=1}^{4}P(J_{0i})dJ_{0i} \log \left\{ 2\cosh \{ \beta(J_{01}+J_{02}+J_{03}+J_{04})\right\} = 2 \log 2,\\
\end{eqnarray}
\end{widetext}
where $P(J_{ij})$ is the Gaussian distribution function with the mean $J_0$
and the variance $J=1$. The numerical manipulation of this equation gives
the location of the multicritical point for the Gaussian Ising model on the
square lattice as $J_0^{(1)} = 1.021564$.

\subsection{Other Approximations for the Square Lattice}

\begin{figure}[tb]
\begin{center}
\includegraphics[width=75mm]{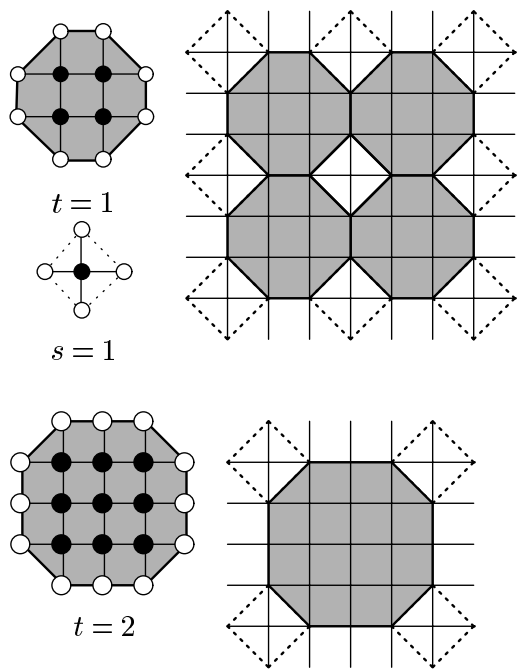}
\end{center}
\caption{{\protect\small Other approximations by use of two clusters on the
square lattice. The small lattice $t=1$ includes $12$ bonds and $t=2$ has $24
$ bonds.}}
\label{other}
\end{figure}

We restrict ourselves to the case of the square lattice and consider other
types of the improvement. The key of the improvement is the cluster
reflecting the shape of the square lattice. We consider here another type of
approximations by dividing the square lattice into two clusters, which can
cover the whole of the lattice, as in Fig. \ref{other}. Then we can reduce
the duality relation (\ref{PF}) to, 
\begin{eqnarray}
& & Z^{(s,t)}_n(x^{(s)}_0,x^{(s)}_1,\cdots;x^{(t)}_0,x^{(t)}_1,\cdots) 
\notag \\
& & \quad =
Z^{(s,t)}_n(x^{*(s)}_0,x^{*(s)}_1,\cdots;x^{*(t)}_0,x^{*(t)}_1,\cdots),
\end{eqnarray}
where $Z^{(s,t)}_n$ is the reduced partition function after the summation
over internal sites on two clusters, and $x^{(s)}$ and $x^{(t)}$ are the
edge Boltzmann factor including many-body interactions after the summation
denoted by $s$ and $t$, respectively. The reduced partition function is
regarded as a multi-variable function of two types of arguments $x^{(s)}$
and $x^{(t)}$. We extract the principal Boltzmann factors and assume that a
single equation gives the location of the multicritical points, 
\begin{equation}
x_0^{(s)}x_0^{(t)} = x_0^{*(s)}x_0^{*(t)}.
\end{equation}
From this equation, we estimate the location of the multicritical point. For 
$s=1$ and $t=1$ small lattices, we obtain $p^{(1,1)}_c = 0.890794$ and, for $%
s=1$ and $t=2$ small lattices, $p^{(1,2)}_c = 0.890813$. If the cluster
includes many bonds, we expect that the improved technique can approach the
answer. It is then considered that the multicritical point is located at $%
p_c = 0.890813$. At least, we find that both of the estimations indicate a
higher value $p_c \approx 0.8908$ than $p^{(0)}_c \approx 0.8900$, and the
precise location of the multicritical point would be $p_c \approx 0.8908$. 

\subsection{First Approximation for the Potts Spin Glass}

As another application of the improved technique, let us consider the Potts
spin glass on the square lattice defined by the Hamiltonian, 
\begin{equation}
H = - J \sum_{\langle ij \rangle} \delta\left( \phi_{ij} + l_{ij}\right),
\label{4PottsSG}
\end{equation}
where $\phi_{ij} \equiv \phi_i - \phi_j$ expresses the difference between
adjacent Potts spins taking an integer value between $0$ and $q-1$. The
quantity $l_{ij}$ is the random variable following the distribution function
given as 
\begin{eqnarray}
P(l_{ij}) = \left\{ 
\begin{array}{ll}
1-(q-1)p & (l_{ij} = 0) \\ 
p & (l_{ij} \neq 0) \\ 
& 
\end{array}
\right\} = \frac{\mathrm{e}^{K_p\delta(l_{ij})}}{\mathrm{e}^{K_p}+q-1},
\label{Pdis}
\end{eqnarray}
where $\mathrm{e}^{K_p} \equiv \{1-(q-1)p\}/p$. This Potts spin glass also
has the gauge symmetry. For the Potts spin variables and random variables,
we define the gauge transformation as, 
\begin{eqnarray}
\phi_i &\to& \phi_i + s_i, \\
\l _{ij} &\to& \l _{ij} - (s_i - s_j).
\end{eqnarray}
Here $s_i$ takes an integer between $0$ and $q-1$. Therefore we can
establish the Nishimori line by setting $\beta J = K_p$, where the internal
energy can be calculated exactly and the specific heat can be bounded \cite%
{4HNPotts}. The edge Boltzmann factor for the Potts spin glass is given as, 
\begin{equation}
x(\phi_{ij}) = \mathrm{e}^{\beta J \delta(\phi_{ij}+l_{ij})},
\end{equation}
and the dual one is, 
\begin{equation}
x^*(\phi_{ij}) = \frac{v}{\sqrt{q}}\left\{ \mathrm{e}^{\mathrm{i}\frac{2\pi}{%
q}l_{ij}\phi_{ij}} + \frac{q}{v} \delta(\phi_{ij})\right\},
\end{equation}
where $v \equiv \mathrm{e}^{\beta J} -1$.

We thus give the conventional conjecture as follows, \cite{NN,MNN}, 
\begin{eqnarray}
& & - \left\{ 1 - (q-1) p \right\} \log \left\{ 1-(q-1)p \right\} - (q-1) p
\log p  \notag \\
& & \quad = \frac{1}{2}\log q.
\end{eqnarray}
The solutions are obtained as $p_c = 0.079731$ for $q=3$, $p_c = 0.063097$
for $q=4$, and $p_c = 0.052467$ for $q=5$.

We here estimate the location of the multicritical point for the Potts spin
glass on the square lattice by use of the $s=1$ small lattice. The partition
functions on the cluster as in Eq. (\ref{6IC1}) are given as, 
\begin{eqnarray}
& & Z^{(1)}(\beta;\{l_{ij}\})  \notag \\
& & \quad = \left[ \overline{\sum_{\{\phi_i\}}} \prod^{\mathrm{part}%
}_{\langle ij \rangle} \mathrm{e}^{\beta J \delta(\phi_{ij}+l_{ij})} \right]%
_{\mathrm{av}}, \\
& & Z^{*(1)}(\beta;\{l_{ij}\})  \notag \\
& & \quad = \left[ \overline{\sum_{\{\phi_i\}}} \prod^{\mathrm{part}%
}_{\langle ij \rangle} \frac{v}{\sqrt{q}}\left\{ \mathrm{e}^{\mathrm{i}\frac{%
2\pi}{q}l_{ij}\phi_{ij}} + \frac{q}{v} \delta(\phi_{ij})\right\} \right]_{%
\mathrm{av}}.
\end{eqnarray}
We can carry out the summation over $\phi_0$ at the center of the $s=1$
cluster as in Fig. \ref{fig29r} as, 
\begin{eqnarray}
& & Z^{(1)}(\beta, \{l_{ij}\})  \notag \\
& & \quad = \sum^{q-1}_{\phi_0=0}\mathrm{e}^{\beta J\left\{\delta(\phi_0 +
l_{01}) + \delta(\phi_0 + l_{02}) + \delta(\phi_0 + l_{03})+ \delta(\phi_0 +
l_{04})\right\}}  \notag \\
& & \qquad = q + 4v + v^2\sum_{i \neq j}\delta(l_{0i},l_{0j})  \notag \\
& & \quad \qquad + v^3\sum_{i\neq j \neq k}\delta(l_{0i},l_{0j},l_{0k}) +
v^4\delta(l_{01},l_{02},l_{03},l_{04}),  \notag \\
\end{eqnarray}
where $i \neq j$ means the summation over different pairs among four bonds,
and $i \neq j \neq k$ expresses the summation over all combinations of
different three bonds among four bonds. In addition, the dual principal
Boltzmann factor is given as, 
\begin{eqnarray}
& & Z^{*(1)}(\beta, \{l_{ij}\})  \notag \\
& & \quad = \left( \frac{v}{\sqrt{q}}\right)^4\sum^{q-1}_{\phi_0=0}
\prod^4_{i=1} \left\{ \mathrm{e}^{\mathrm{i}\frac{2\pi}{q}l_{0i}\phi_0} +%
\frac{q}{v}\delta(\phi_0)\right\}  \notag \\
& & \qquad = \frac{v^4}{q^2}\left\{ \left(1+\frac{q}{v}\right)^4 - 1 + q
\delta\left(\sum^4_{i=1}l_{0i}\right)\right\}.
\end{eqnarray}
From these quantities, we rewrite Eq. (\ref{6IC1}) as follows, 
\begin{widetext}
\begin{eqnarray}
\left[ \log \left\{ \frac{\left(q+v\right)^4 - v^4 + q v^4
\delta\left(\sum^4_{i=1}l_{0i}\right)}{q + 4v + v^2\displaystyle\sum_{i \neq
j}\delta(l_{0i},l_{0j}) + v^3\displaystyle\sum_{i\neq j \neq
k}\delta(l_{0i},l_{0j},l_{0k}) + v^4\delta(l_{01},l_{02},l_{03},l_{04})}%
\right\}\right]_{\mathrm{av}} = 2 \log q,  \notag \\
\end{eqnarray}
\end{widetext}
where the configurational average for the random variables $\{l_{0i}\}$ on
the four bonds follows the distribution function (\ref{Pdis}). We obtain $%
p^{(1)}_c = 0.0791462$ for $q=3$, $p^{(1)}_c = 0.0626157$ for $q=4$, and $%
p^{(1)}_c = 0.0520578$ for $q=5$. These results are shown in Table \ref%
{6ICTable} for comparison with those by the conventional conjecture and the
existing result by a numerical estimation \cite{4Potts}.

\subsection{First Approximation for the Triangular Lattice}

\begin{figure}[tb]
\begin{flushleft}
\includegraphics[width=75mm]{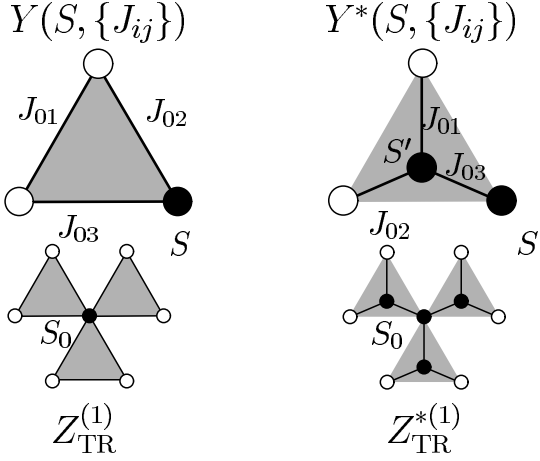}
\end{flushleft}
\caption{{\protect\small Relationships between $Y$ and $Z_{\mathrm{TR}}^{(1)}
$, and between $Y^*$ and $Z_{\mathrm{TR}}^{*(1)}$}. The spin $S^{\prime }$
at the center of the star shape is of the summation for the star-triangle
transformation.}
\label{fig32}
\end{figure}

We show the explicit calculation of the first approximation by the improved
technique for the $\pm J$ Ising model on the triangular lattice. We
consider the cluster labeled by $s=1$ with three up-pointing triangles as in
Fig. \ref{fig30}. In this case, it is convenient to define the following
quantities, 
\begin{eqnarray}
& & Y(S, \{J_{ij}\})  \notag \\
& & \quad = \mathrm{e}^{\beta(J_{01}+J_{02}S+J_{03}S)}, \\
& & Y^{*}(S, \{J_{ij}\})  \notag \\
& & \quad = \frac{1}{4} \sum_{S^{\prime }=\pm 1} \prod_{i=1}^2 \left( 
\mathrm{e}^{\beta J_{0i}} + S^{\prime }\mathrm{e}^{-\beta J_{0i}}\right)
\left( \mathrm{e}^{\beta J_{03}} + S^{\prime }S\mathrm{e}^{-\beta
J_{03}}\right)  \notag \\
& & \qquad = \frac{1}{4} \left\{\prod_{i=1}^{2}\left(2\cosh \beta
J_{0i}\right)\left( \mathrm{e}^{\beta J_{03}} + S\mathrm{e}^{-\beta
J_{03}}\right) \right.  \notag \\
& & \left. \quad \qquad + \prod_{i=1}^{2} \left(2\sinh \beta J_{0i}\right)
\left( \mathrm{e}^{\beta J_{03}} -S\mathrm{e}^{-\beta J_{03}}\right)\right\},
\end{eqnarray}
where the locations of the spins $S^{\prime }$, $S$, and the sets of $%
\{J_{0i}\}$ are described in Fig. \ref{fig32}. The summation over $S^{\prime
}$ corresponds to the star-triangle transformation. 
We calculate two partition functions for the cluster as, by use of two
quantities $Y$ and $Y^*$, 
\begin{eqnarray}
& & Z_{\mathrm{TR}}^{(1)}(\beta, \{J_{ij}\})  \notag \\
& & \quad = \sum_{S_0=\pm 1}\prod^3_{k=1}Y(S_0, \{J_{k,ij}\})  \notag \\
& & \qquad = 2\mathrm{e}^{\beta \sum^{3}_{k=1}J_{k,01}} \cosh \{ \beta
\sum^{3}_{k=1}\left(J_{k,02}+J_{k,03}\right)\}, \\
& & Z_{\mathrm{TR}}^{*(1)}(\beta, \{J_{ij}\})  \notag \\
& & \quad = \sum_{S_0=\pm 1}\prod^3_{k=1}Y^*(S_0, \{J_{k,ij}\})  \notag \\
& & \qquad = 2^3 \left\{\prod_{k=1}^{3}\left(\prod_{i=1}^3 \cosh \beta
J_{k,0i} + \prod_{i=1}^3 \sinh \beta J_{k,0i}\right) \right.  \notag \\
& & \left.\quad \qquad + \prod_{k=1}^{3} \left(\prod_{i=1}^2 \cosh \beta
J_{k,0i} \sinh \beta J_{k,03} \right.\right.  \notag \\
& & \left. \left. \qquad \qquad + \prod_{i=1}^2 \sinh \beta J_{k,0i} \cosh
\beta J_{k,03} \right) \right\}.
\end{eqnarray}
Substituting these quantities into Eq. (\ref{6ICTR}), we can obtain the
first approximation by the improved technique on the triangular lattice.
The result is given as $p_c^{(1)} = 0.835957$, which is also listed in Table %
\ref{6ICTable} with the one $p^{(2)}_c = 0.835985$ by another approximation
by the numerical manipulation of the $s=2$ cluster in Fig. \ref{fig30}. Each
estimation gives the precise locations of the multicritical point on the
hexagonal lattice $p^{(1)}_c = 0.932611$ and $p^{(2)}_c = 0.932593$, by the
relation for the mutually dual pair \cite{TSN}, 
\begin{equation}
H(p_{\mathrm{TR}}) + H(p_{\mathrm{HEX}}) = 1,  \label{4MDConjecture2}
\end{equation}
where $H(p)$ is the binary entropy defined by 
\begin{equation}
H(p) = -p\log_2 p -(1-p)\log_2(1-p).
\end{equation}

Even on the triangular lattice, we can consider the Potts spin glass \cite%
{OPotts}, and apply the improved technique. In this case, we can describe,
by the improved technique, a more precise line consisting of the
multicritical points on a two-dimensional plane of couplings representing
two-body interactions and three-body interactions. 
\begin{table}[tbp]
\begin{center}
\begin{tabular}{lll}
\hline
Type & Conjecture & Numerical result \\ \hline
SQ $\pm J$ & $p^{(0)}_c = 0.889972$ \cite{NN,MNN} & $0.8905(5)$ \cite{Aarao}
\\ 
& $p^{(1)}_c = 0.890725$ & $0.8906(2)$ \cite{Hone,Picco} \\ 
& $p^{(1,1)}_c = 0.890794$ & $0.8907(2)$ \cite{Merz} \\ 
& $p^{(2)}_c = 0.890822$ & $0.8894(9)$ \cite{Ito} \\ 
& $p^{(1,2)}_c = 0.890813$ & $0.8900(5)$ \cite{Queiroz} \\ 
&  & $0.89081(7)$ \cite{Hasen} \\ 
SQ Gaussian & $J^{(0)}_0 = 1.021770$ \cite{NN,MNN} & $1.02098(4)$ \cite%
{Picco} \\ 
& $J^{(1)}_0 = 1.021564$ &  \\ 
TR $\pm J$ & $p^{(0)}_c = 0.835806$ \cite{NO} & $0.8355(5)$ \cite{Queiroz}
\\ 
& $p^{(1)}_c = 0.835956$ &  \\ 
& $p^{(2)}_c = 0.835985$ &  \\ 
HEX $\pm J$ & $p^{(0)}_c = 0.932704$ \cite{NO} & $0.9325(5)$ \cite{Queiroz}
\\ 
& $p^{(1)}_c = 0.932611$ &  \\ 
& $p^{(2)}_c = 0.932593$ &  \\ 
SQ Potts($q=3$) & $p^{(0)}_c = 0.079731$ \cite{NN,MNN} & 0.079-0.080 \cite%
{4Potts} \\ 
& $p^{(1)}_c = 0.079146$ &  \\ 
SQ Potts($q=4$) & $p^{(0)}_c = 0.063097$ \cite{MNN} & -- \\ 
& $p^{(1)}_c = 0.062616$ &  \\ 
SQ Potts($q=5$) & $p^{(0)}_c = 0.052467$ \cite{MNN} & -- \\ 
& $p^{(1)}_c = 0.052058$ &  \\ \hline
\end{tabular}%
\end{center}
\caption{{\protect\small Several predictions for the location of the
multicritical point by the improved technique. SQ denotes the square
lattice, TR expresses the triangular lattice, and HEX means the hexagonal
lattice. }}
\label{6ICTable}
\end{table}

\section{Performance of Improvement}

\subsection{Multicritical Point}

We here discuss the performance of the improved technique.

At first, we remark the predictions for the location of the multicritical
point of the $\pm J$ Ising model on the square lattice. All of the results
by the improved technique for the $\pm J$ Ising model on the square lattice
indicate a higher value about $p_c \approx 0.8908$ than $p_c^{(0)} \approx
0.8900$ by the conventional conjecture. As the size of the cluster
increases, the prediction of $p_c$ converges to some value about $p_c
\approx 0.8908$. We need the precision to the fourth digit to conclude the
conflict between $p_c \approx 0.8900$\cite{Ito,Queiroz} and $p_c \approx
0.8908$\cite{Aarao,Hone,Picco,Merz,Hasen}. For this purpose, the improved
technique gives a satisfactory answer that the multicritical point is
located at $p_c \approx 0.8908$. We cannot completely deny the possibility
that the multicritical point locates at $p_c \approx 0.8900$ as estimated in
other studies \cite{Ito,Queiroz}, because the improved technique does not
give the exact solution. However the following discussions support our
conclusion $p_c \approx 0.8908$ from a different point of view.

\subsection{Phase Diagram}

\begin{figure}[tb]
\begin{center}
\includegraphics[width=75mm]{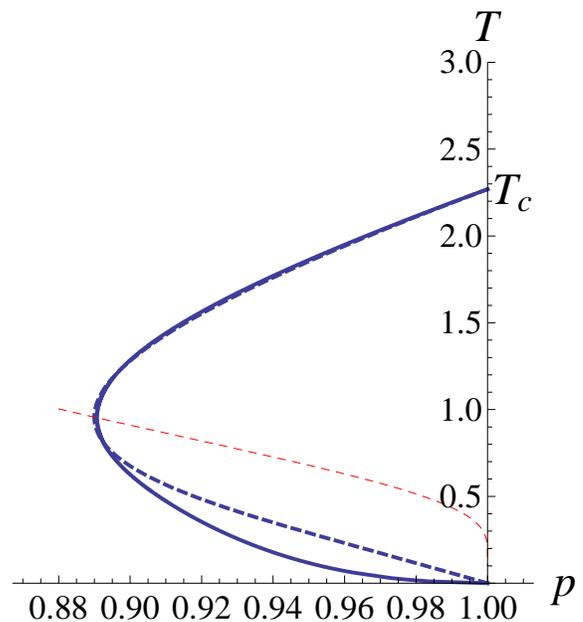}
\end{center}
\caption{{\protect\small (Color online) Phase diagram of the $\pm J$ Ising model on the
square lattice by the $s=1$ improved technique. The vertical axis is the
temperature, and the horizontal axis is the probability for $J_{ij}=J>0$ of
the $\pm J$ Ising model. The thick dashed line (blue) is by the conventional
conjecture and the thick solid line (blue) is by the improved technique. The thin
dashed line (red) is the Nishimori line.}}
\label{improvedphase}
\end{figure}

\begin{figure}[tb]
\begin{center}
\includegraphics[width=75mm]{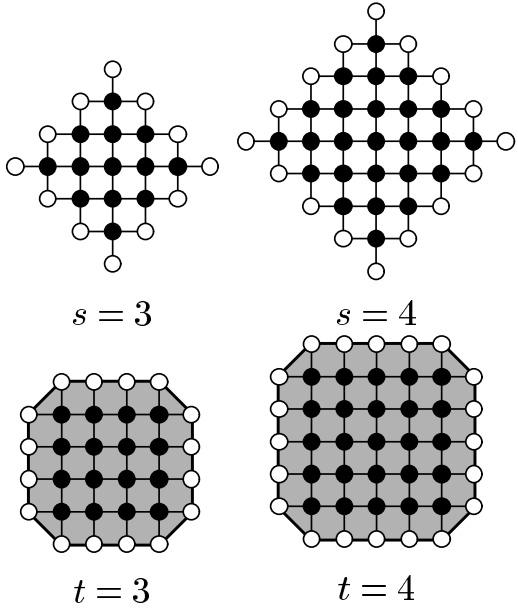}
\end{center}
\caption{{\protect\small Several approximations for estimations of the slope
at $T_c$. The numbers of bonds in each pattern are $36~(s=3)$, $40~(t=3)$, $%
64~(s=4)$, and $60~(t=4)$.}}
\label{fig31}
\end{figure}

The phase boundary can be predicted by the improved technique without the
restriction of the Nishimori-line condition, similarly to the conventional
conjecture. Unfortunately the improved technique fails again to derive the
precise phase boundary especially under the Nishimori line similarly to the
case of the hierarchical lattice \cite{ONB} as in Fig. \ref{improvedphase}. 
Nevertheless we find an improvement of estimations of the slope of the phase
boundary at the critical point $T_c$ of the non-random Ising model. We
concentrate on the estimations of the slope and show the results below by
use of not only $(s=1)$, $(s=2)$, $(s=1,t=1)$, and $(s=1,t=2)$ clusters but
also several ones as in Fig. \ref{fig31}. 
The computing time of the order $O(2^{N_B^{(s)}})$ is needed in general for
the configurational average over $\{J_{ij}\}$ in evaluation of the improved
technique. However, the configurational average becomes much simpler, when
we consider a calculation only around $T_c$ to estimate the slope at $T_c$,
at which at most a single bond becomes antiferromagnetic. Therefore we can
deal with further approximations only to estimate the value of the slope at $%
T_c$ by various clusters. The obtained values are listed in Table \ref%
{TcSlopeTable}. 
\begin{table}[tbp]
\begin{center}
\begin{tabular}{ccl}
\hline
Type & Number & Value of slope \\ \hline
$s=0$ & $1$ & $3.41421$ \\ \hline
$s=1$ & $4$ & $3.33658$ \\ 
$s=2$ & $16$ & $3.31272$ \\ 
$s=3$ & $32$ & $3.29352$ \\ 
$s=4$ & $64$ & $3.28161$ \\ \hline
exact &  & $3.20911$ \cite{Domany1} \\ \hline
\end{tabular}
\begin{tabular}{ccl}
\hline
Type & Number & Value of slope \\ \hline
$s=0$ & $1$ & $3.41421$ \\ \hline
$s=1,t=1$ & $4+12$ & $3.31225$ \\ 
$s=1,t=2$ & $4+24$ & $3.29414$ \\ 
$s=1,t=3$ & $4+40$ & $3.28170$ \\ 
$s=1,t=4$ & $4+60$ & $3.27287$ \\ \hline
exact &  & $3.20911$ \cite{Domany1} \\ \hline
\end{tabular}%
\end{center}
\caption{{\protect\small Slope at the critical point $T_c$ for the $\pm J$
Ising model on the square lattice. The top table gives the results by the
clusters with many cross shapes denoted by $s$. The bottom table shows those
by two clusters represented by $(s,t)$. For comparison, we write the result
by the conventional conjecture denoted by $s=0$. }}
\label{TcSlopeTable}
\end{table}
Two types of the approximations, by use of one clusters and by dividing the
square lattice into two clusters, give different values but, in any cases,
the increase of the size of the cluster shows convergence to the exact
solution $3.20911$ of the slope at $T_c$ by Domany \cite{Domany1}. Therefore
it is considered that the improved technique gives a systematic way to
derive the precise locations of the critical points in the region especially
above the Nishimori line.

\subsection{Verticality and Reentrance}

We also examine the shape of the phase boundary at the multicritical point
of the $\pm J$ Ising model. The following observation also supports the
validity of the improved technique. Taking the derivative by $\beta$ of Eq.
(\ref{6IC2}), we obtain the expression of the slope of the phase boundary at
the multicritical point, here denoted by $\beta^{(s)}_c$, 
\begin{equation}
\left.\frac{d\beta_p}{d\beta}\right|_{\beta=\beta^{(s)}_c} = - \frac{%
\displaystyle\sum_{\{J_{ij}\}}\left(\frac{1}{2^{N_{\mathrm{D}}}}\frac{d Z^*}{%
d \beta} - \frac{1}{2^{N_{\mathrm{s}}}}\frac{d Z}{d \beta} \right)}{%
\displaystyle\sum_{\{J_{ij}\}}\left(\frac{1}{2^{N_{\mathrm{D}}}}\frac{dZ^*}{%
d\beta_p}\log Z^* - \frac{1}{2^{N_{\mathrm{s}}}}\frac{dZ}{d\beta_p}\log
Z\right)},
\end{equation}
where we omit the approximation type $s$, and the arguments of $Z$ and $Z^*$
for simplicity. One can find the quantities in the numerator in the
right-hand side of this equation being equal to the exact internal energy on
the Nishimori line \cite{HN81,Rev3}, 
\begin{equation}
\sum_{\{J_{ij}\}}\frac{1}{2^{N_{\mathrm{D}}}}\frac{d Z^*}{d \beta} =
\sum_{\{J_{ij}\}} \frac{1}{2^{N_{\mathrm{s}}}}\frac{d Z}{d \beta} = N_B
\tanh \beta_p J.
\end{equation}
Hence the slope of the phase boundary at the multicritical point should be
vertical. In other words, the multicritical point is located at a minimum
value $p_c$ on the phase boundary predicted by the improved technique. This
statement is satisfied for any clusters. Therefore, even if we consider the
infinite size of the small lattice in which we can expect to obtain the
exact answer, the verticality at the multicritical point holds. The improved
technique can give the consistent phase boundary with the predicted by the
gauge symmetry \cite{HN81,Rev3}.

The second derivative yields a non-zero value of $d^2 \beta_p/d \beta^2$,
which is proportional to the difference of the specific heat between $\pm J$
Ising models on two clusters, 
\begin{eqnarray}
& & \left.\frac{d^2\beta_p}{d\beta^2}\right|_{\beta=\beta^{(s)}_c}  \notag \\
& & \quad = - \frac{\displaystyle\sum_{\{J_{ij}\}}\left\{\frac{1}{2^{N_{%
\mathrm{D}}}}\frac{1}{Z^*}\left(\frac{d Z^*}{d \beta}\right)^2 - \frac{1}{%
2^{N_{\mathrm{s}}}}\frac{1}{Z}\left(\frac{d Z}{d\beta}\right)^2 \right\}}{%
\displaystyle\sum_{\{J_{ij}\}}\left(\frac{1}{2^{N_{\mathrm{D}}}}\frac{dZ^*}{%
d\beta_p}\log Z^* - \frac{1}{2^{N_{\mathrm{s}}}}\frac{dZ}{d\beta_p}\log
Z\right)},  \notag \\
\end{eqnarray}
We do not have the exact value of the specific heat on the Nishimori line
though we know an upper bound \cite{HN81,Rev3}. We cannot completely
determine the shape of the phase boundary only by the improved technique.
However we remark that the estimated values of the second derivative become
lower, if the cluster under consideration become larger as $0.956729~(s=0)$, 
$0.753892~(s=1)$, and $0.737262~(s=2)$. These positive values indicate that
the phase boundary predicted by the improved technique become reentrant or
vertical. The possibility of the phase boundary is indeed limited into
whether vertical or reentrant as rigorously shown by the gauge-symmetry
argument \cite{HN81,Rev3}.

\subsection{Other Random Spin Systems}

If the improvement affects the predictions not only of the multicritical
point but also for other critical points, we can apply the improved
technique to random spin systems without gauge symmetry. The absence of the
Nishimori line on the phase diagram does not permit us to rewrite Eq. (\ref%
{6IC1}) as Eq. (\ref{6IC2}), which can give a relation between two entropies
of the distribution of frustration. However the previous discussions by the
duality for the cluster are applicable to various random spin models. We
then give the critical points by the following equation even for random spin
systems without the Nishimori line, 
\begin{eqnarray}
& & -\beta \left(\left[ F_{\mathrm{D}}^{(s)} \right]_{\mathrm{av}} - \left[
F^{(s)} \right]_{\mathrm{av}}\right)  \notag \\
& & \quad = \left(\frac{N^{(s)}_B}{2} - N^{(s)}_s + 1 \right) \log2,
\label{6IC3}
\end{eqnarray}
where $\left[F^{(s)}\right]_{\mathrm{av}}$ is the free energy on the
cluster, and $\left[F_{\mathrm{D}}^{(s)}\right]_{\mathrm{av}}$ represents
that on the dual cluster. Therefore our task to analytically derive the
critical points in random spin systems is to estimate the difference between
the two free energies on the cluster and its dual one.

\begin{figure}[tb]
\begin{center}
\includegraphics[width=75mm]{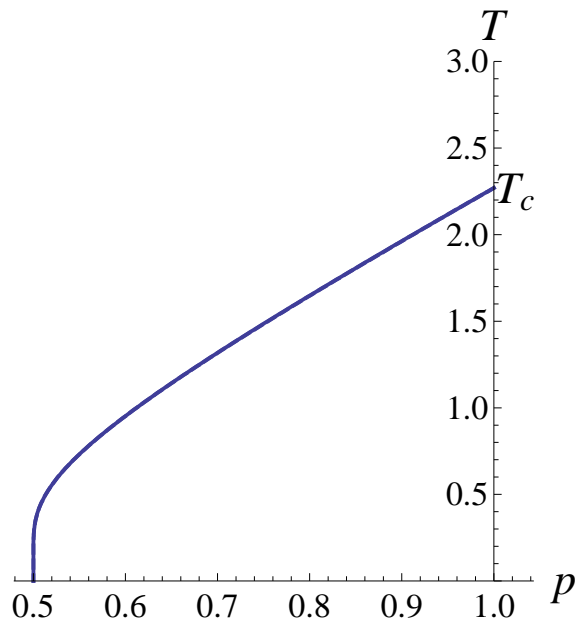}
\end{center}
\caption{{\protect\small (Color online) Phase diagram of the bond-diluted Ising model on
the square lattice by the $s=1$ improved technique. The vertical axis is
the temperature, and the horizontal axis is the probability for $J_{ij}=J>0$.%
}}
\label{phasedilute}
\end{figure}

We apply the improved technique to the bond-diluted Ising model by the
evaluation of Eq. (\ref{6IC3}) by use of the distribution function, 
\begin{equation}
P(J_{ij}) = p \delta(J_{ij}-J) + (1-p)\delta(J_{ij}).
\end{equation}
The predicted phase boundary in Fig. \ref{phasedilute} is not drastically
different from the one by the conventional conjecture \cite{HNcon2}. 
However we can find a significant difference by investigation of the values
of the slope at $T_c$ similarly to the case of the $\pm J$ Ising model. We
show the results for the slope at $T_c$ for the bond-diluted Ising model in
Table \ref{TcSlopeTable2}. 
\begin{table}[tbp]
\begin{center}
\begin{tabular}{ccl}
\hline
Type & Number & Value of slope \\ \hline
$s=0$ & $1$ & $1.34254$ \\ \hline
$s=1$ & $4$ & $1.33780$ \\ 
$s=2$ & $16$ & $1.33626$ \\ 
$s=3$ & $32$ & $1.33500$ \\ 
$s=4$ & $64$ & $1.33420$ \\ \hline
exact &  & $1.32926$ \cite{Domany2} \\ \hline
\end{tabular}
\begin{tabular}{ccl}
\hline
Type & Number & Value of slope \\ \hline
$s=0$ & $1$ & $1.34254$ \\ \hline
$s=1,t=1$ & $4+12$ & $1.33623$ \\ 
$s=1,t=2$ & $4+24$ & $1.33504$ \\ 
$s=1,t=3$ & $4+40$ & $1.33421$ \\ 
$s=1,t=4$ & $4+60$ & $1.33362$ \\ \hline
exact &  & $1.32926$ \cite{Domany2} \\ \hline
\end{tabular}%
\end{center}
\caption{{\protect\small Slope at $T_c$ for the bond-diluted Ising model on
the square lattice. For comparison, we write the result by the conventional
conjecture denoted by $s=0$. }}
\label{TcSlopeTable2}
\end{table}
The estimated values for the square lattice shows convergence to the exact
solution $1.32926$ \cite{Domany2}, similarly to the case for the $\pm J$
Ising model.

In addition, we remark that the improved technique works very well for the
critical points of the bond-diluted $q$-state Potts model, and $q$-state
Villain model \cite{Okabe}. {}From these points of view, we conclude that
the improved technique is also a systematic approach leading to the precise
locations of the critical points in broader classes of the random spin
systems.

\section{Conclusion}

We proposed an improved technique applicable to the square, triangular,
hexagonal lattices, and derived the precise locations of the multicritical
points for the $\pm J$ Ising model, the Gaussian Ising model, and the Potts
spin glass on the square lattice, as well as the $\pm J$ Ising model on the
triangular lattice and the hexagonal lattice. 
This improved technique is still approximation for the location of the multicritical point.
However we can enhance the precision of the approximation by the summation over spins in the cluster taken from the considered lattice, if we need the precise location of the critical points in a random spin system.
This would open a way to analytically derive the location of the critical points
in random spin systems with very high precision. Unfortunately, in the
low-temperature region under the Nishimori line, the improved technique
cannot give satisfactory answers yet. We solve this problem in the
low-temperature region under the Nishimori line, and have to examine the
validity of some hypotheses on the improved technique.

In this paper, we restrict ourselves to the random spin systems in
two-dimensional systems. However we can apply the duality to other
dimensional systems. For example, the duality can transform the random-bond
Ising model on the three-dimensional cubic lattice into the random-plaquette
gauge model on the three-dimensional cubic lattice. The random-plaquette
gauge model is an attractive one in terms of the quantum toric code \cite%
{DKLP,WHP}. An accuracy threshold to correct error of the quantum toric code
corresponds to the location of the multicritical point on the
random-plaquette gauge model with the random couplings following the $\pm J$
distribution function on the three-dimensional cubic lattice. The
conventional conjecture relates this threshold with the location of the
multicritical point of the $\pm J$ Ising model on the three-dimensional
cubic lattice \cite{TN}. The improved technique also cannot directly derive
such an accuracy threshold, but can make more precise relationship between
the locations of the multicritical points on the random-bond Ising model and
the random-plaquette gauge model.

As another direction of studies in the future, we should clarify the
physical meaning of the equation consisting of the entropy of the
distribution of frustration, which determines the location of the
multicritical point. 
\begin{acknowledgments}
The author greatly acknowledges the fruitful discussion with Prof. H.
Nishimori, Prof. A. N. Berker and Dr. M. Hinczewski, and sincerely thanks
Prof. Y. Okabe for sending unpublished numerical data of the bond-diluted
Potts model and the bond-diluted Villain model, and Prof. M. Picco, who
informed him of a his recent result of the multicritical point on the square
lattice. He would like to also thank Prof. S. L. A. de Queiroz for a
valuable comment and Dr. K. Takahashi for reading the manuscript and giving
stimulating comments. This work was partially supported by the Ministry of
Education, Science, Sports and Culture, Grant-in-Aid for Young Scientists
(B) No. 20740218, and for scientific Research on the Priority Area
\textquotedblleft Deepening and Expansion of Statistical Mechanical
Informatics (DEX-SMI)\textquotedblright , and by CREST, JST.
\end{acknowledgments}



\end{document}